\documentclass{ws-p8-50x6-00}

\def\bfp{{\bf p}}  
\begin{document}

\title{Hadrons in the Nuclear Medium- Role of Light Front Nuclear Theory}

\author{Gerald A. Miller}

\address{Department of Physics, University of Washington\\
  Seattle, WA 98195-1560\\
E-mail: miller@phys.washington.edu}


\maketitle

\abstracts{
The problem of understanding the nuclear effects observed in lepton-nucleus
deep-inelastic-scattering (the EMC effect) is still with us.
Standard nuclear models (those using only
hadronic degrees of freedom)
are not able to account for  the EMC effect.  Thus it is necessary 
 to understand how the 
nuclear medium modifies quark wave functions  in the  nucleus. 
Possibilities for such modifications,  represented by the 
 quark meson coupling model, and the suppression of point-like-configurations
are discussed, and  methods to
experimentally choose between these are reviewed.
}

\section{Introduction}
When the organizers asked me to give a talk entitled
``Hadrons in the Nuclear Medium'' I thought about what the title
might mean. Since 
 the nuclear mass $M_A\approx[N M_n+Z M_p] (1-0.01)$, and nucleons are hadrons,
 maybe the title
 should be 
``Hadrons are   the Nuclear Medium!''. On the other hand, the modern paradigm
for the strong interaction is QCD and 
 QCD is a theory of quarks and gluons. Maybe the title should be
``Are Hadrons the Nuclear Medium?''. We have known, since the discovery of the 
 EMC effect in 1982, that the structure functions
 measured in deep inelastic scattering from nuclear targets
 are not those of free  nucleons. So one theme of this talk is to try to
 understand, 
 interpret  and use  the EMC effect. Despite the age of this effect, no
 consensus has been reached regarding its interpretation, importance and
 implications.

 The second theme of this talk arises from the kinematic variables used to
 describe the data. The Bjorken $x$ variable:
$x={Q^2/ 2M\nu}$ is, in the parton model, a ratio of
quark to target momenta $ {p^+_q/ P^+_A}$, where the
superscript $+$ refers the plus-component of the four-momentum vector.
This in turn can be written as: $
({p^+_q/ p^+_N})\;({p^+_N/P^+_A}),$ so that one needs to know
how often a nucleon has a given value of plus-momentum, $p^+_N$.
Conventional nuclear wave functions are not expressed in terms of this
variable, so one needs to derive nuclear wave functions which are expressed in
terms of plus-momenta. Therefore, I assert that 
 light front nuclear theory is needed. 
\section{\bf Outline}

I turn towards a more detailed outline. The first part of the talk is concerned
with what I call the 
  ``Return of the EMC effect''. This is the statement that conventional nuclear
  physics does not explain the EMC effect.
The physics here is subtle, so I believe that 
  some formal development  involving  the construction of nuclear
  wave functions using light front nuclear theory is needed. I try to
  answer the 
  simple queries:
``light front theory- what is it? why use it?''. A partial answer is
reviewed in Ref.~\cite{Miller:2000kv}.
The  saturation  of infinite
nuclear matter using the mean field approximation is discussed here.
The result is 
that there is
no binding effect which explains the
EMC effect\cite{Birse:hu},\cite{Miller:2001tg}.
This statement is a natural
consequence of the light version of the 
  Hugenholtz-Van Hove Theorem\cite{HvH}: the pressure of a stable system vanishes.
  If one goes beyond the mean-field calculations and includes
  correlations by using a Hamiltonian which involves only nucleon degrees of
  freedom, there is again  no binding effect. Furthermore, I'll argue that
  using mesons along with nucleons probably won't allow a   description of
   all the
  relevant  data. 

It is therefore reasonable, proper and necessary to examine the subject
of how the internal structure of a
   nucleon is  modified  by the nucleus to which it is bound.
   To see how the
   medium modifies the wave function of nucleon, a 
   particular model of the free proton wave function\cite{Schlumpf:ce} is
   used. This allows the examination of 
 two different and complementary  ideas:  
the  quark meson coupling model\cite{Guichon:1987jp} and the
 suppression of point-like-configurations\cite{Frankfurt:cv}.

A quick  summary is that the   
 goal is to  explain EMC effect 
 and then  predict new 
experimental consequences. 
 This goal is not attained but is within reach.

\section{Return of the EMC effect}\label{subsec:remc}
It is necessary to use 
 nuclear wave functions in which one of the variables is
the plus-momentum of a nucleon, $p_N^+$. Thus the use of 
 light front quantization, or light-front dynamics, which I now try 
to explain, is necessary.

\subsection{Light  Front Quantization Lite}

Light-front dynamics
is a relativistic many-body dynamics in which fields are quantized at a
``time''=$\tau=x^0+x^3\equiv x^+$. The $\tau$-development operator
 is then given by
$P^0-P^3\equiv P^-$. These equations show the  notation that    
a four-vector $A^\mu$ is expressed in terms of its $\pm$ components
$ A^\pm\equiv A^0\pm A^3.$
One quantizes at $x^+=0$ which is a light-front, hence the name ``light front
dynamics''.
 The canonical spatial variable must be orthogonal to the time variable,
 and this is given by 
$x^-=x^0-x^3$. The canonical momentum is then $P^+=P^0+P^3$. The other
coordinates are as usual ${\bf x}_\perp$ and  ${\bf P}_\perp$.

 The most important  consequence of this is that the
 relation between energy and
momentum of a free particle is given by:
\be p_\mu p^\mu=m^2=p^+p^--p_\perp^2\;\to\;  p^-={p_\perp^2+m^2\over p^+},\ee
 a relativistic formula for the kinetic energy which does not contain
a square root operator. This feature
allows the  separation of  center of mass
and relative coordinates, so that the computed wave functions are frame
independent.

The philosophy, for the beginning of this talk,  is 
 to use a Lagrangian density, ${\cal L}$ which is  
 converted into an energy-momentum tensor $ T^{\mu\nu}$.
 The total-four-momentum operator is defined formally
as 
\be
P^\mu ={1\over 2}\int d^2x_\perp dx^- T^{+\mu}, \ee
with  $P^+$ as the ``momentum'' operator and $P^-$ as the  ``energy''
operator. We then need to
 express $ T^{+\mu}$
in terms of independent variables. In particular, the 
 nucleon, usually described as a four-component spinor, is a  spin 1/2
particle and therefore there are only two  independent degrees of freedom.

I'll start with  the well-known 
 Walecka model  in which ${\cal L}(\phi,V^\mu,N)$
 is expressed in terms of nucleon $N$, vector meson  $V^\mu$, and scalar
 meson $\phi$ degrees of freedom.  The  plan  is to first carry
 out calculations using  the 
 mean field approximation, and then include the effects of $NN$
correlations using other Lagrangians.

\subsection            {\bf Light Front Quantization }

 The mode equation for nucleons in infinite  nuclear matter
nuclear matter is given by
\be\left({\not\! k}- g_v{\not\!V}-(M+g_s\phi)\right)\psi=0,\ee      
within the mean field approximation of the Walecka model. The quantity
of relevance for understanding deep inelastic scattering is the nuclear plus
component of momentum given by\cite{Miller:2001tg} 
\bea P_A^+=\langle A\vert\psi_+^\dagger\left(i\partial^++g_vV^+\right)\psi_+
\vert A\rangle,\label{paplus}\eea
where $\psi_+={1\over 2}\gamma^0\gamma^+\psi$, the
independent component of the nucleon field. Furthermore
\be P_A^-=\langle A\vert \psi_+^\dagger\left({-\partial^2_\perp+(M+g_s\phi)^2
\over i\partial^+}\right)\psi_+\vert A\rangle+m_s^2\phi^2. \ee
In the rest frame we must have $P^+_A=M_A,\;P^-_A=M_A=E_A$, a result not
obvious from the above equations. However, if we 
 minimize $P^-_A$ subject to the constraint that the expectation value of
 $(P^+_A-P^-_A)$ vanish, we indeed get the 
 same $E_A$ as Walecka and more! The more refers to information about the
 plus momenta.

The result
$P^+_A=P^-_A$ means $P^3_A=0,$ which is a statement that
pressure vanishes for a stable system. According to a venerable 1958
theorem by Hugenholtz \& Van Hove, a  vanishing pressure, plus
the definition
that the
nucleon Fermi energy
$E_F\equiv \left( {\partial E_A\over \partial A}\right)_{\rm
volume},$ gives  
$E_F={E_A\over A}={M_A\over A}\equiv \overline{M}$. This  has important
consequences now because we may express the result
(\ref{paplus}) as
\be P^+_A=M_A=A\int dk^+ f_N(k^+)\; k^+.\ee Next we use a dimensionless
variable
$y\equiv {k^+\over\overline{M}}$ to find that
  \be  \int f_N(y)\;y=1, \label{lfsr}\ee which means that nucleons carry all of the plus
  momentum.

\def\be{\begin{equation}}
 \def \ee{\end{equation}}
\def\bea{\begin{eqnarray}}\def\eea{\end{eqnarray}}
\def\eqn {Eq.~(\ref }


%



The relevance of this can be seen by calculating the effects of nucleons
in deep inelastic lepton nucleus scattering using a manifestly
covariant calculation of the handbag diagram. One finds\cite{jm}
\be
{F_{2A}(x_A)\over A}=\int^\infty_{x} dy f_N(y) F_{2N}(x_A/y),\;\ee
where 
\be  f_N(y)=\int {d^4k\over (2\pi)^4} \delta(y-{k^0+k^3\over \overline{M}})
Tr\left[ {\gamma^+\over A}\chi(k,P)\right],\;
x_A=x M/ \overline{M}.\ee
The quantity $\chi(k,P)$ is the nuclear expectation value of the
connected part of the nucleon Green's function. This can easily be
calculated for our light front nuclear wave functions.
The result is  
\bea
f_N(y)&=&4\int {dk^+d^2k_\perp \over(2\pi)^3\rho_B}
\theta\left(k_F^2-k_\perp^2-(k^++g_vV^+-E_F)^2\right)\nonumber\\
&&\times\delta(y-{k^+ +g_vV^+\over\overline{M}})
                                \eea
or
\be
f_N(y)=
{3\over 4} {\overline{M}^3\over k_F^3}\theta(1+k_F/\overline{M}-y)
\theta(y-(1-k_F/\overline{M}))\left[
{k_F^2\over \overline{M}^2}-(1-y)^2\right],
\ee       which obeys
the baryon  $\int dy f_N(y)=1$ and momentum (\ref{lfsr}) sum rules, so that
 nucleons carry all of the plus momentum. 

This is important because 
$f_N(y)$ is narrowly peaked at $y=1$, so that
$ 
{F_{2A}(x_A)}\approx 
AF_{2N}(x_A), $
and  there is 
almost NO  Binding Effect.

One can see this more directly by 
expanding $F_2(x/y)$ in Taylor
series\cite{Frankfurt:1985ui} about $y=1$ to find:
\bea
F_{2N}(x_A)/A
= F_{2N}(x)
+{\epsilon}{F'_{2N}(x)} 
+{k_F^2\over 10 \overline{M}^2}
(2x_AF'_{2N}(x_A)+x_A^2F''_{2N}(x_A)),\label{he}        \eea
where
$ \epsilon=1-\overline{M}/M_N\approx  {16/940}.$
The resulting figure is shown as Fig.~2 of Ref.\cite{Miller:2001tg},
but Eq.~(\ref{he})  shows clearly that 
${F_{2N}(x_A)/ F_{2N}(x)}$ is too large.

Similar calculations can now be done for finite nuclei
\cite{Smith:2002ci}.
Although being able to do these calculations is a major technical
achievement (according to me), the  results also show a huge disagreement with
experiment.

\def\be{\begin{equation}}
 \def \ee{\end{equation}}   
\def\bea{\begin{eqnarray}}\def\eea{\end{eqnarray}}
\subsection{ Beyond the Mean Field Approximation}
Suppose one 
assumes  nucleons are the only degrees of freedom in the Hamiltonian, so that
\be H=\sum_i t_i \;+\sum_{i<j} v_{ij}\;+\sum_{i<j<k}v_{ijk}+\cdots.\ee
Any correct
 solution gives must be consistent with the Hugenholtz Van Hove Theorem
$P^+=P^-=M_A=P^+_N,$ so that 
once more one finds \be\int dy\; f_N(y)\;y=1.\ee 

One may again expand make an expansion about $y=1$ to find that 
\be
{F_{2N}(x_A)/A} = 
F_{2N}(x)+{\langle\epsilon\rangle\over \overline{M}}{F'_{2N}(x)\over F_{2N}(x) }
+\gamma \left(2x_AF'_{2N}(x_A)+ x_A^2F''_{2N}(x_A)\right),\ee
where
$\gamma=\int\;dy\;f_N(y) (y-1)^2.$
Again there is NO binding effect, and the results are even worse, in comparison
with experiment, than before.
Using a more elaborate many-body calculation with a Hamiltonian involving only
nucleons can not explain the EMC effect.

\subsection{\bf Nucleons $N$ and  Mesons $m$}
The best version of the conventional approach is to explicitly  include
the effects of mesons. Then one may compute the nuclear expectation 
value of  $T^{++}$ as
\be P_A^+=P_N^++P_m^+=M_A, \qquad\epsilon\equiv {P_m^+\over M_A},\ee
so that one would find
\be\int dy\;f_N(y)\;=1-\epsilon.\ee
Many authors can reproduce the 
 EMC binding effect using $\epsilon\approx 0.07$,
but $\epsilon\approx 0.07$ corresponds to a BIG enhancement of the
nuclear sea. Some time ago theorists suggested that the nuclear 
 Drell-Yan process could be used to
 disentangle the EMC effect\cite{Bickerstaff:ax}, but 
 data from E772 at Fermilab\cite{Alde:im}
 showed no enhancement, and no nuclear effects.
This finding  was termed a 
 ``Crisis in Nuclear Theory''\cite{missing}.

\subsection{Summary of Return of the  EMC Effect}
Conventional nuclear theory is unable to provide an explanation of the
EMC effect. Mean field theory gives no EMC effect. Any theory involving
using only nucleons as degrees of freedom gives no EMC effect. Including
the effects of mesons explicitly (and not buried in the nucleon-nucleon
potential) can provide an explanation or description of the EMC effect, but
seems to cause a disagreement with the Drell-Yan data.  Thus it is entirely
legitimate, correct and proper to take very seriously the proposition that
the structure of the nucleon is modified by its presence in the nuclear medium.
\section
{ Non-Standard Nuclear Physics }

The idea that modification of nucleon properties has important experimental
consequences can be termed as non-standard nuclear physics. The main goal I
wish to discuss here is that of first
finding some model (or models) that reproduce
both the EMC effect and the  nuclear Drell-Yan data, and then predict
other experimental 
consequences. There are
 many ideas in the literature. We shall take up only two here.
The first involves 
 nucleon wave function modification by the  nuclear mean field. This is the
  quark meson coupling model introduced by Guichon\cite{Guichon:1987jp}
and applied to understanding the 
  EMC effect by  Saito and Thomas\cite{Saito:yw}. 
  The nucleus is bound by the effects of mesons which are exchanged
  between quarks in different nucleons, so that the
  wave function of a quark in a
  bound nucleon is
  modified.

 The second idea involves the suppression of point like configurations (PLC),
introduced by 
 Frankfurt and Strikman\cite{Frankfurt:cv}
 as a consequence of the effects of color neutrality
of nucleons. The idea is that the nucleon consists of an infinite number of
 configurations of different sizes, each having different interactions with
 the nuclear medium. There are configurations with anti-quarks and gluons which
are large and blob-like (BLC). These are influenced by the attractive force
provided by other nucleons. There are also rarer configurations consisting of
only three quarks which are close to each other. These PLC do not interact
because the effects of gluons emitted by such a color-singlet
configuration are canceled.
The energy differences between the BLC and the rarer PLC are increased by
the nuclear medium, so the probability of the PLC are decreased. This is the
suppression we speak of.
The validity of this  idea was
checked and confirmed
by   Frank, Jennings and Miller\cite{Frank:1995pv}.

Our attitude is the the quark meson coupling model and the suppression of
PLC are different reasonable hypotheses for nuclear modifications of nucleon
wave functions which should be taken seriously.
\subsection {Light Front Model of Proton}

The basic idea is to put a nucleon in the medium and study how it responds to
the external forces provided by the other nucleons.
We need a relativistic model, and a convenient one is that of
Schlumpf\cite{Schlumpf:ce}, which was recently exploited
by us\cite{Frank:1995pv,Miller:2002qb}.
The model wave function is written in terms of
light-cone variables and can be written schematically
as
\bea 
\Psi(p_i) \;=\;u(p_1) u(p_2) u(p_3)
 \psi(p_1,p_2,p_3),\eea
in which $p_i$ represent 
space (cm), spin and isospin variables, and in which the 
{  $u$} { are conventional Dirac spinors}.
The function 
$\psi$ depends on
 spin and isospin and includes a spatially symmetric function
 $ \Phi(M_0^2)$. The quantity
 $M_0^2$ is the square of the mass of a non-interacting 
system of three quarks which plays the role that the square of spatial
three momentum would in an ordinary wave function.
Thus:
\bea
M_0^2=\sum_{i=1,3} {p_{i\perp}^2+m^2\over p_i^+}\qquad      
\Phi(M_0^2)={N'\over (M^2_0+\beta^2)^{\gamma}}\;,
\eea
 The wave function is now specified. It is 
 expressed in terms of relative variables and is a
boost invariant light front wave function. 

The first application is 
 how the electromagnetic form factors are modified in the medium,
 so we need to consider the model's version of the form factors of a free
 nucleon.
  These are obtained by sandwiching the wave functions around the current
 operator  $J^+\sim \gamma^+$.
 The evaluation of this Dirac operator is
 simplified by making a unitary transformation to represent the wave function
 $\Psi$ in terms of light front spinors which have the nice property:
$\bar u_L(\bfp'\lambda')\gamma^+u_L(\bfp\lambda)=2p^+\delta_{\lambda\lambda'}.$
This allows us to interpret the results in an analytic fashion. The
coefficients
of the unitary transformation are known as the Melosh transformation, and one
example is given by 
\bea
\bar u_L(\bfp_3,\lambda_3)
u(\bfp_3, s_3)= \langle \lambda_3\vert
\left[ {m+(1-\eta)M_0+i{ \sigma}\cdot({\bf n}\times {\bf p}_3)\over
\sqrt{(m+(1-\eta)M_0)^2+p_{3\perp}^2}}\right]\vert s_3\rangle.
\eea
The basic idea is that the
spin flip term given by $i{ \sigma}\cdot({\bf n}\times {\bf p}_3)$
is as large as the non-spin flip term given by $m+(1-\eta)M_0$. For large
momentum
transfer, $Q$, {\it each} of these is proportional to $Q$. The form factor  $F_1$
depends on the non-spin-flip term:  $F_1\sim Q\cdots$, and $F_2$ depends on the
spin-flip term, so $QF_2\sim
Q\cdots,$ as well. Thus
 the ratio
${QF_2/ F_1}$ is approximately  constant for sufficiently large $Q$.
The results are shown in
Fig.~1 and are in
good agreement with the recent
exciting data\cite{Jones:1999rz},\cite{Gayou:2001qd} . 
This means we have a reasonable model nucleon to put in the nuclear medium.

\begin{figure}
\unitlength1cm
\begin{picture}(11,9)(0,-10.5)
\includegraphics{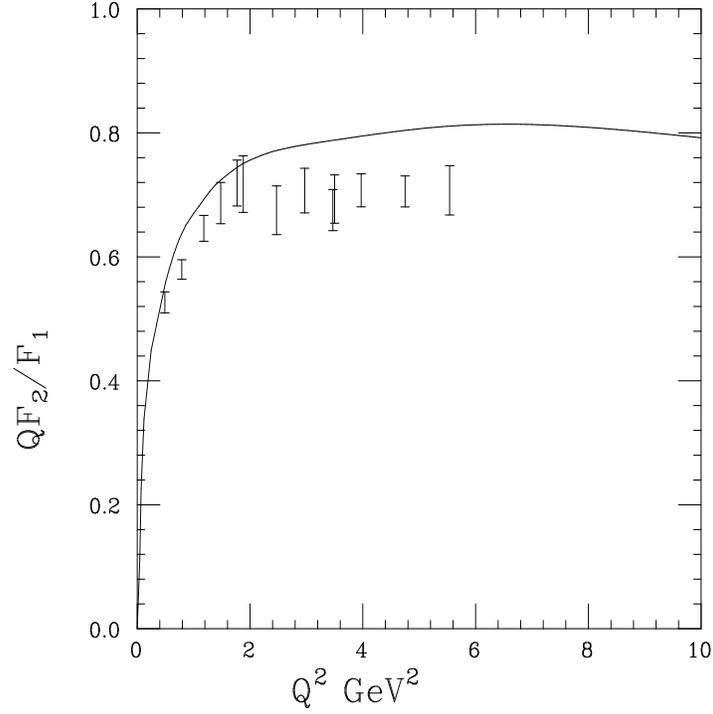}
\end{picture}
\label{fig:ratio}
\caption{Calculation of Refs.~$^{15,16}$ 
  the data
  are from  Ref.~$^{17}$ for 
$2\le Q^2\le 3.5$ GeV$^2$  and from  Ref.~$^{18}$ 
$3.5\le Q^2\le 5.5\; {\rm GeV}^2$} 
\end{figure}
\subsection {Medium modifications of nuclear form factors}

Let's start with the quark meson coupling model
QMC. We approximate  that the 
 nuclear scalar $\sigma$  and vector potentials are
constant over the volume of the nucleon. Then the modification is simply
expressed as
$m\to m-\sigma$, with the average scalar field
experienced by a quark is given by $\sigma\approx 40 \;{\rm MeV}$. Thus we
simply
reduce the quark  mass used in the previous calculation by 40 MeV. The results
are shown in Figs.~2 
and 3. 

\begin{figure}
\unitlength1cm
\begin{picture}(11,9)(0,-10.5)
\includegraphics{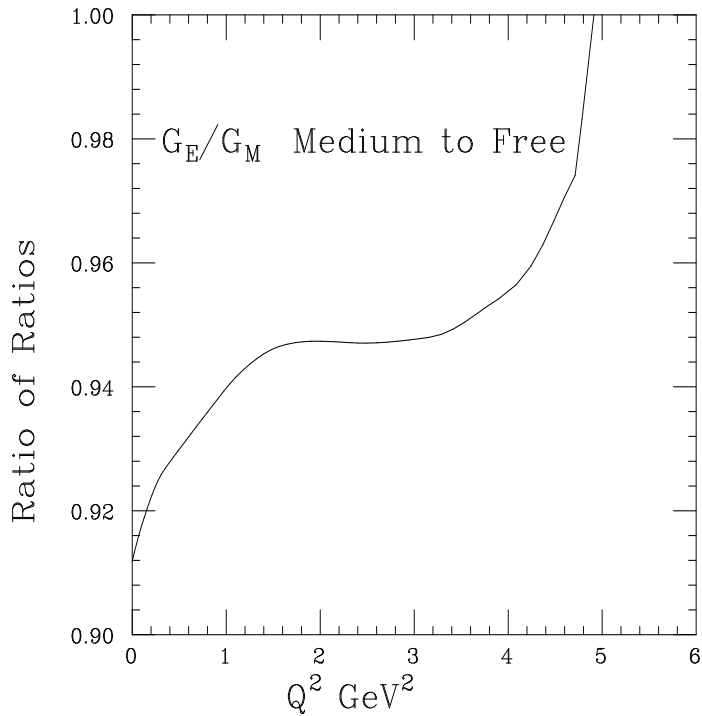}
\end{picture}
\caption {Medium modifications of the ratio $G_E/G_M$, QMC}
\label{fig:mratio}
\end{figure}
\begin{figure}
\unitlength1cm
\begin{picture} (1,9)(0,-8.5)
\includegraphics{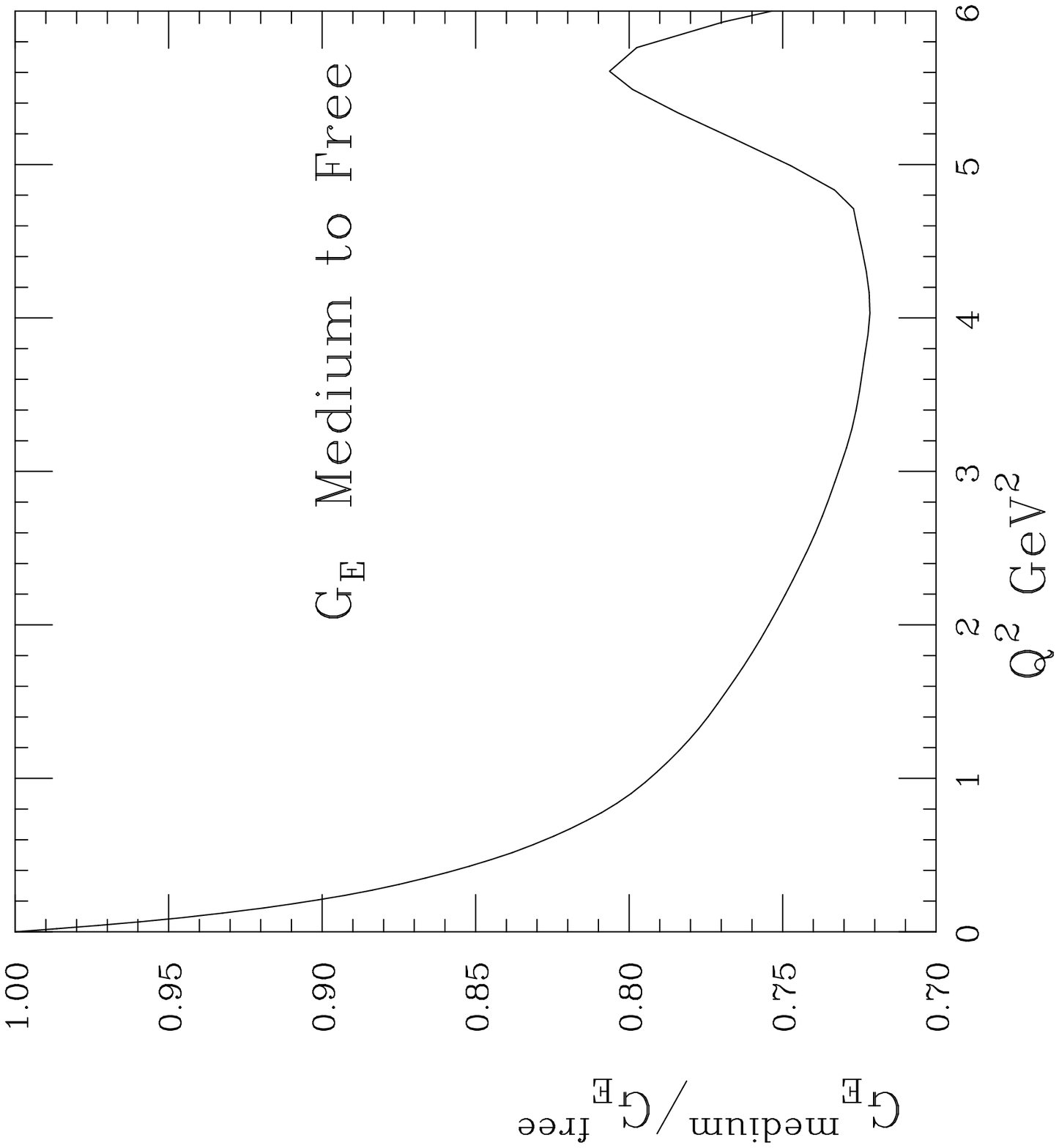}
\end{picture}
\label{fig:eratio}
\caption {Medium modifications of the ratio $G_E/G_M$, QMC}
\end{figure}
These results seem to give huge effects, but one must understand that these
are form factors evaluated
 in the nuclear ground state.
The only attempts to observe such effects have involved using the
$(e,e'p)$ reaction. Thus only the proton in the initial state is  modified
in the manner described here.
Furthermore, the reaction occurs at the nuclear surface where the
$\sigma$ field is falling off towards zero. When such realities are included,
the results would be  similar to the  theory of Ref.~\cite{Lu:1998tn}, and 
similar to the $^4$He data of  Dieterich {\it et al.}\cite{Dieterich:sj} and the
$^{16}$O data Malov {\it et al.}\cite{Malov:rh}.
For $^4$He, the actual effects are
about four times smaller than shown here.
\subsection{\bf Suppression of PLC}

The idea here is that the interaction of a bound
nucleon  with a nucleus   depends
on the distances between the quarks. The relevant  operator is 
\be
r^2\equiv \sum_{i<j} (\vec{r}_i-\vec{r}_j)^2,\ee
so that the nuclear mean field $U$ is given by 
\be U(r,R)=U_0 \rho(R) {r^2\over \langle N\vert r^2 \vert N\rangle},\ee
which vanishes for PLC.
Then, in first-order perturbation theory, the modification of the expectation
value  of
any operator ${\cal O}$ is given by
\be\delta \langle {\cal O}\rangle = 2{U_0\rho(R)\over \Delta E}
\left({\langle N\vert {\cal O}\; r^2\vert N\rangle
\over \langle N\vert r^2 \vert N\rangle}
-\langle N\vert{\cal O}\;\vert N\rangle\right).\ee
The resulting  form factors  
are shown in Figs.~10,11 of
Ref.~\cite{Frank:1995pv}.  The effects of the medium are smaller here than for
the QMC. But  there are substantial modifications of the
valence quark distribution functions, as shown in Fig.~16 of
Ref.~\cite{Frank:1995pv}. These  are relevant for understanding the EMC effect.
\subsection{Tagged structure functions}
There are  many ideas available to explain EMC effect. More experiments are
needed to select the correct ones. Here I only have room to discuss one
possibility\cite{Melnitchouk:1996vp} in which one studies the 
reaction: $ e'D\to e+N +X$ for the kinematics of deep-inelastic
scattering.
The ratio 
\be G(x_1,x_2)\equiv{\sigma(x_1,k^+,{\bf k}_\perp,Q^2)\over
\sigma(x_1,k^+,{\bf k}_\perp,Q^2)}
={F_{2N}^{\rm mod}(x_1/(2-k^+/M),{\bf k}_\perp,Q^2)\over
F_{2N}^{\rm mod}(x_2/(2-k^+/M),{\bf k}_\perp,Q^2)
}\ee is extremely   sensitive to the different models, in which
models giving the same DIS have differences of more than $ 50\% $. 
This is shown in Fig.~6
of Ref.~\cite{Melnitchouk:1996vp}.   
\section{Summary}
The use of the standard,
conventional meson-nucleon dynamics of nuclear physics is not
able to explain the nuclear deep-inelastic and Drell-Yan data.
The logic behind this can be understood using the Hugenholtz-van Hove
theorem\cite{HvH} which states that the stability (vanishing of pressure)
causes
the energy of the single particle state at the Fermi surface to be 
$M_A/A\approx 0.99 M_N$.
In light front language,
 the
vanishing pressure is achieved by obtaining $P^+=P^-=M_A$.
But $P^+=\int dk^+f_N(k^+)k^+$.
This, combined with the relatively small value of the Fermi momentum
(narrow width of $f_N(k^+)$)
means that
the probability $f_N(k^+)$
for a nucleon to have a given value of $k^+$ must be narrowly
peaked about $k^+=0.99 M_N\approx M_N.$
Thus the effects of nuclear binding and Fermi motion play only a very limited
role in the nuclear structure function, and the  resulting  function  
must  very close to the one of a free nucleon unless some quark-gluon effects
are included. This means that  
some
non-standard explanation involving quark-gluon degrees of freedom is
necessary.
Many contending non-standard ideas available. Testing these models,
selecting the right ones, and ultimately determining the dynamical significance
depends on having new high accuracy experiments at large momentum transfer and
energy.

\section*{Acknowledgments} This work is partially supported by the U.S. DOE.

\end{document}